# 3D time-domain beam mapping for studying nonlinear dynamics in multimode optical fibers

Y. Leventoux,[1] G. Granger,[1] K. Krupa,[2] A. Tonello,[1] G. Millot,[3] M. Ferraro,[4] F. Mangini,[5] M. Zitelli,[4] S. Wabnitz,[4,6] S. Février,[1] and V. Couderc,[1,*]

[1] *Université de Limoges, XLIM, UMR CNRS 7252, 123 Av. A. Thomas, 87060 Limoges, France*
[2] *Institute of Physical Chemistry, Polish Academy of Sciences, ul. Kasprzaka 44/52, 01-224 Warsaw, Poland*
[3] *Université Bourgogne Franche-Comté, ICB, UMR CNRS 6303, 9 Av. A. Savary, 21078 Dijon, France*
[4] *DIET, Sapienza University of Rome, Via Eudossiana 18, 00184 Rome, Italy*
[5] *DII, University of Brescia, Via Branze 38, 25123 Brescia, Italy*
[6] *Novosibirsk State University, Pirogova 1, Novosibirsk 630090, Russia*
*Corresponding author: vincent.couderc@unilim.fr*

**The characterization of the complex spatiotemporal dynamics of optical beam propagation in nonlinear multimode fibers requires the development of advanced measurement methods, capable of capturing the real-time evolution of beam images. We present a new space-time mapping technique, permitting the direct detection, with picosecond temporal resolution, of the intensity from repetitive laser pulses over a grid of spatial samples from a magnified image of the output beam. By using this time-resolved mapping, we provide the first unambiguous experimental observation of instantaneous intrapulse nonlinear coupling processes among the modes of a graded index fiber.**

Over the past two decades, research on multimode optical fibers (MMFs) had an impressive comeback in several research areas. This was initially motivated by their use for spatial-division-multiplexing (SDM) in optical communication systems, to overcome the capacity crunch of singlemode fiber transmissions. Soon thereafter, nonlinear properties of MMFs attracted a resurgence of interest for fundamental research as well [1]. MMFs add new dimensions to optical wave propagation, and provide an excellent testbed to study nonlinear light complexity and collective wave dynamics. Among others, we may mention multimode solitons [2-4], geometric parametric instabilities (GPI) [5,6], Kerr beam self-cleaning [7,8], classical wave condensation [9,10], and spatiotemporal (ST) mode-locking [11].

The study of ST complexity inherent to nonlinear propagation in MMFs represents a major challenge. Recently, new theoretical and computational methods have been developed, to facilitate numerical simulations [12-14]. In order to fully characterize and understand the ST beam dynamics in MMFs, it is also crucial to develop new advanced experimental techniques. This key issue has been addressed only quite recently. Gabolde and Trebino introduced a technique to measure the complete ST intensity and phase of a single pulse, by means of simultaneous multiple holograms, each at a different color [15-16]. Wang et al. [17] developed of a new real-time multidimensional measurement technique, based on spatiotemporal-spectral (STS) compressed ultrafast photography (CUP). This permitted to measure the non-repeatable, shot-to-shot spatial evolution of 3D dissipative solitons, as well as ST variations of multi-pulse 3D solitons in each cavity round-trip. Although very powerful, the STS-CUP technique is relatively complex, and it requires the use of a streak camera, which is not standard laboratory equipment. A mode-resolved STS analysis with a streak camera was also developed in Ref. [18], in order to measure the spectral broadening induced by cross-phase modulation in a few-mode fiber.

In this Letter, we present a new ST measurement method, based on the use of a real-time digital oscilloscope with deep memory. Our method is best suited for long laser pulses, which can be revealed by direct detection. In particular, our method allows to measure pulse-to-pulse, repeatable spatiotemporal phenomena with picosecond temporal resolution. This permits us to provide the first experimental evidence of ST nonlinear power exchange between the fundamental and higher-order modes (HOMs) of a graded-index (GRIN) MMF. Nonlinear mode coupling leading to Kerr beam self-cleaning is characterized in real-time across the laser pulse. This permits to show how a speckled beam transforms into a bell-shaped beam across different levels of the instantaneous pulse intensity, in agreement with a theoretical description of the nonlinear mode-coupling process [7]. An indirect proof of these power-dependent temporal dynamics was previously obtained by measuring the temporal evolution when spatially sampled in a limited zone at the beam center [19].

In our experiments, we used a microchip laser at 1064 nm, delivering 1.5 ns pulses (measured at $1/e^2$ of their maximum) with a repetition rate of 27 kHz. The linearly polarized Gaussian laser beam was injected into a GRIN MMF. The beam diameter at the input face of the fiber was 40 μm (measured at $1/e^2$). We used a GRIN MMF with 25 μm core radius, a refractive index difference of 0.015, and a numerical aperture of 0.2. We applied our novel multidimensional scanning technique to the study of the ST dynamics of a multimode beam in the process of Kerr self-cleaning. Additionally, we used a silicon camera, in order to perform a reference measurement, as it was done in previous beam self-cleaning experiments. Note that a silicon camera integrates the beam intensity over a temporal window that is longer than the nanosecond laser pulse duration.

Our ST mapping technique is schematically illustrated in Fig.1. The magnified image of the output of the GRIN MMF is scanned with a singlemode fiber (SMF) which is moved in the transverse plane. The temporal profiles of the light transmitted by the SMF are detected by a high-speed 45-GHz photodiode, and stored, for each position of the SMF, in the memory of a 70-GHz real-time oscilloscope. This allows for a temporal resolution of 8 ps, which is much shorter than the laser pulse duration in our experiments. With a suitable post-processing, we segment the large array stored in the oscilloscope according to the SMF scanning positions. The obtained 3D matrix then makes it possible to reconstruct, at each temporal point, the spatial profile of the output beam. We used a motorized 3-axis stage for automatic scanning through the SMF segment. Moreover, to avoid pulse-to-pulse fluctuations (for instance related to the timing jitter of the Q-switched laser), a small fraction of the laser beam was exploited to trig and self-reference the oscilloscope for each laser pulse via a second silicon photodiode.

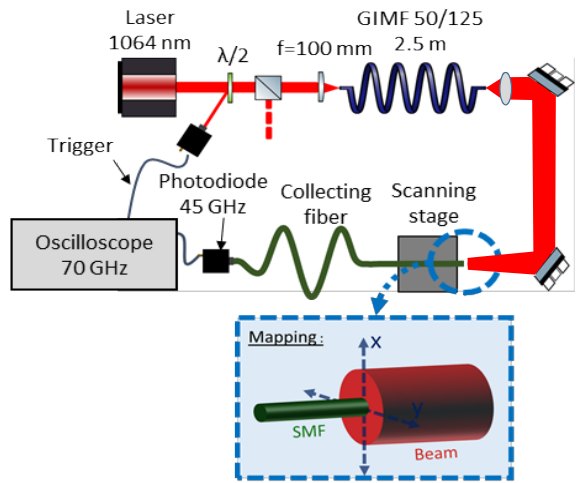

Fig. 1. Experimental set-up for spatiotemporal beam mapping.

In order to validate our novel characterization technique, we first performed the measurement of the input laser beam, before its injection into the fiber. The obtained 3D measurement, including two spatial (x, y) and one temporal (t) dimension, is illustrated in Fig.2. For the reference, the two side insets (a) and (b) illustrate the comparison with measurements performed with a fast oscilloscope, and using a multimode pigtail (which produces integration in the transverse spatial domain) or a slow-response silicon camera (leading to integration in the time domain), respectively. As we can see from Fig.2, the results of our ST mapping are in excellent agreement with standard characterization techniques: the projection on the spatial (x,y) plane fits well with the time-averaged measurement of the camera (see inset (b) of Fig.2). The spatially-averaged temporal profile measured by connecting the photodiode with a multimode pigtail also agrees well with the temporal shape obtained from our new method, once that spatial integration is applied (see inset (a) of Fig.2).

As a second step, by using our ST technique, we measured the temporal evolution, upon input power, at the output of a 2.5 m long GRIN MMF. This step led us to reproduce the measurements reported in Ref.[19]: here we only scan the area occupied by the fundamental mode.

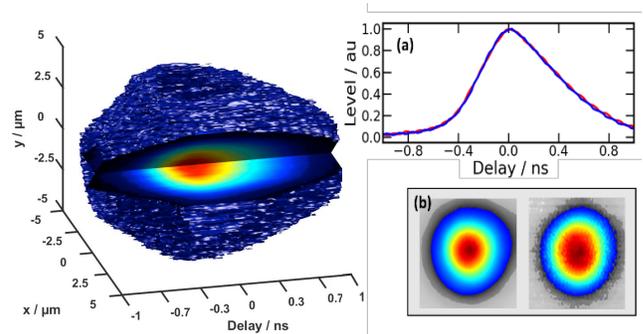

Fig. 2. 3D experimental measurement of the input laser beam before its coupling into the MMF. Inset (a): Comparison between the spatially averaged temporal profile of a single pulse recorded with a fast oscilloscope (red curve) and the temporal profile reconstructed from our 3D measurement, followed by numerical integration over all pixels (blue curve); Inset (b): Spatial pattern registered by a silicon camera (left) compared with the integrated spatial profile obtained from the 3D data array, numerically integrated in time domain (right).

The results are presented in panel (a) of Fig. 3: as it can be seen, we can clearly repeat with the present ST technique the observation of the recurrent behavior of temporal reshaping (see Fig.2 of Ref. [19]), which is associated to the process of coherent nonlinear coupling between the fundamental mode and HOMs (leading to pulse breaking, as highlighted by the dashed circles). Note that such quasi-periodic evolution of pulse breaking is no longer visible when, for each sample of time, we average over the whole transverse domain (see panel (b) of Fig.3). In both cases (i.e., panel (a) and (b) of Fig.3), we can clearly see the local depletion of the pump caused by stimulated Raman scattering (SRS) for pump powers above 24 kW.

Let us consider now the application of our full 3D ST technique to characterize beam self-cleaning in the GRIN MMF. As discussed in Ref.[7], spatial beam cleaning is a two-step process, involving quasi-phase-matched intermodal four-wave-mixing (IFWM), followed by nonlinear nonreciprocity of spatial mode-coupling. As numerically demonstrated initially, the Kerr-induced refractive index grating generated by the periodic oscillation of the beam intensity, or spatial self-imaging, allows for the quasi-phase-matching of IFWM, leading to a periodic energy exchange among the modes. In a second step, which occurs for higher input powers, the self-phase modulation (SPM) cumulated by the fundamental mode stops the periodic back-conversion to HOMs. As a result, the fundamental mode retains indefinitely the accumulated energy, because of the resulting nonlinear nonreciprocity. For reference, the Kerr beam self-cleaning is presented in insets (c)-(f) of Fig.3 in the same way as it has been reported in most of previous publications [7,19], namely by showing near-field images taken by a camera with long response time, as a function of the peak power. The sequence of panels (c-f) illustrates the evolution from a speckled beam at low powers, towards a bell-shaped beam at the center surrounded by a weak, HOM background at high powers.

As reported in Fig. 4, we applied our 3D ST method to perform a series of 3D measurements, for five different input powers. Here we used a 5-m long GRIN fiber span. Our experiments provide the first experimental proof of the theoretically predicted nonlinear periodic coupling between the fundamental and the HOMs. In the linear regime of beam propagation, as well known, a speckled spatial pattern is observed at the fiber output, which results from linear interference between the multiple excited guided modes.

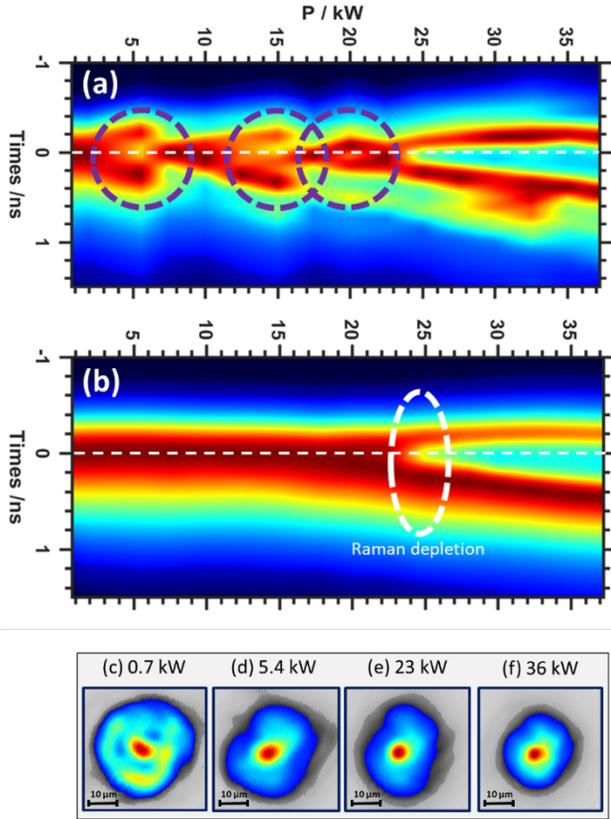

Fig. 3. Evolution of the temporal pulse profile at 1064 nm, vs input peak power, as obtained: (a) from the integration over the area occupied by fundamental fiber mode, or (b) from the integration over all pixels constituting the output image. Insets (c-f): Near-field patterns recorded by a silicon camera at selected power levels in panels (a).

Indeed, as it can be clearly seen in panel (a) of Fig.4, at each point in time across the pulse, only speckled patterns are obtained. Panels (b) and (c) of Fig.4 display the same 3D measurement, but repeated for an input peak power of 1.4 kW and 7.4 kW, respectively. These values are immediately below and above the power threshold for Kerr beam self-cleaning for the chosen GRIN fiber. We may observe that the initial multimode intensity pattern evolves across the pulse. At the peak of the pulse in Fig.4(b) the spatial shape is similar to a $LP_{11}$ mode. This corresponds to a transient multimode state, towards a self-cleaned state, which could be observed for longer fiber lengths. In contrast, pulse edges in Fig.4(b) keep their multimode structure, because of their low instantaneous power. By further increasing the input energy (up to 7.4 kW, see Fig.4(c)), a periodic nonlinear power exchange between the fundamental and HOMs is observed, as power grows across the duration of the pulse. This corresponds to the first step of the nonlinear process, which leads to Kerr beam self-cleaning (i.e., the power reversible nonlinear coupling stage). The presence of a dominant fundamental mode is clearly visible at different time samples inside the pulse shape, corresponding to different instantaneous power levels. Next, panel (d) of Fig.4 shows the 3D ST measurements, performed for a 19 kW peak power. As can be seen, the central part of the pulse is now strongly depleted by SRS. As a result, the remaining part of the pulse temporal envelope at 1064 nm is composed of two consecutive pulses separated by more than 400 ps.

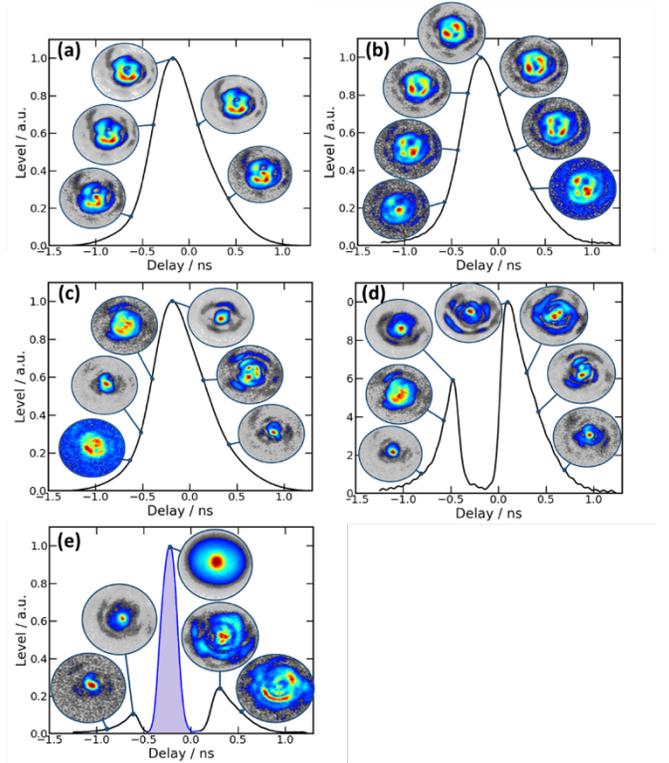

Fig. 4. 2D reconstructed representation of the spatial output pattern at selected instants of time across the pulse duration for different pump powers: (a) 0.75 kW, (b) 1.4 kW, (c) 7.4 kW, (d) 19 kW and (e) 38 kW. The violet pulse of the panel (e) is the converted pulse at the Stokes Raman wavelength (1118 nm).

The periodic evolution of the spatial pattern across the pulse doublet, with a dominant contribution from the fundamental mode, is also observed in this case. At 19 kW of input peak power level is well above the threshold for the second step of the self-cleaning process (where the fundamental mode is decoupled from HOMs via the SPM-induced phase-mismatch which leads to nonlinear nonreciprocity). Thus, SRS-induced pump depletion is the dominant effect. Note that the SRS threshold is lower in Fig.4 than in Fig.2 due to the longer fiber length.

The last curve represented in Fig.4(e), obtained for 38 kW of input peak power, shows the remaining part of the depleted pump pulse and the central pulse part converted at the first Raman Stokes wavelengths (1118 nm - violet color), illustrating the additional Raman-induced beam clean-up in the Stokes pulse.

In order to highlight the multidimensional pulse structuration, we present in Fig. 5 our 3D experimental results of beam self-cleaning, but under a different graphical form. Here we show the measured ST patterns of the output beam, for different input peak power values. Note that, for input powers above the threshold for Raman-induced beam clean-up, a strong spatiotemporal reshaping is perfectly visible. This results in the formation of an isolated, shortened pulse of less than 250 ps of duration, in the time slot that was initially filled by the central part of the input pulse (see Fig.5(e)). Here we may observe a clear competition between Kerr and Raman beam cleanup processes. As a result of the simultaneous energy transfer in the fundamental mode, the SRS power threshold is decreased with respect to the specular propagation.

Our experimental observations are fairly well reproduced by numerical simulations. The numerical model includes, for both polarization components of the electric field envelope, the effect of diffraction, dispersion, and waveguide contribution. The pulse duration is 5 ps, the peak Intensity is 5 GW/cm². Moreover, a sequence of random weak deformations of the fiber, with a maximum extension of 0.1μm, are applied with a coarse step of 5 mm. In Fig.6 we provide an example of our numerical results, showing the 3D ST structuration of the pulse that occurs in the self-cleaning process after 1.5 m of propagation (see also the video of Supplementary visualization 1). The simulation does not include the effect of SRS.

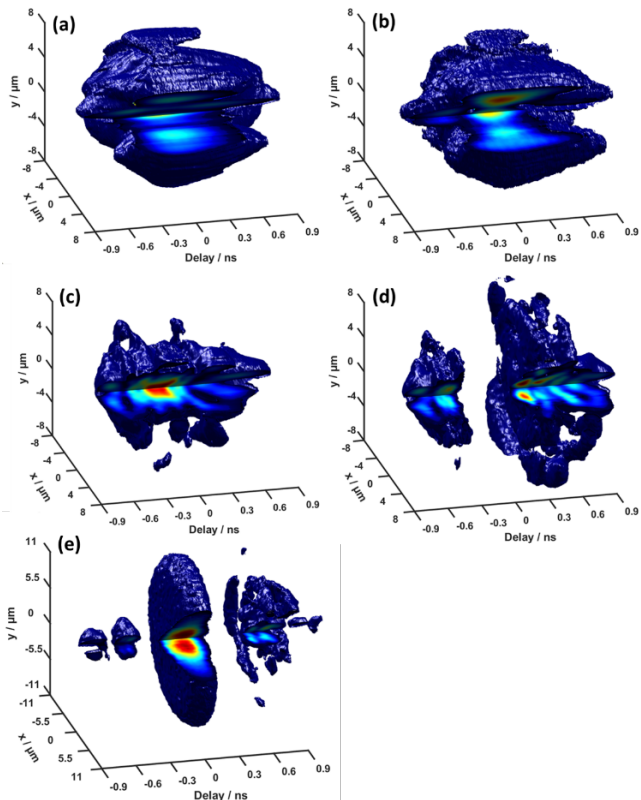

Fig. 5. 3D spatiotemporal patterns of the output beam, vs. input peak power: (a) 0.75 kW, (b) 1.4 kW, (c) 7.4 kW, (d) 19 kW and (e) 38 kW.

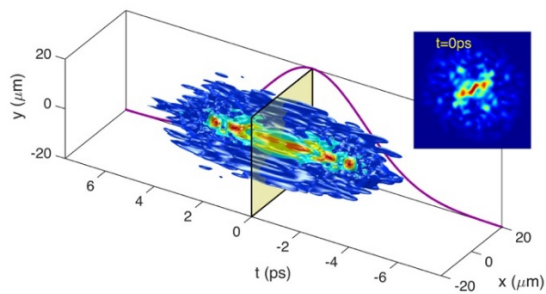

Fig. 6. Numerically calculated ST distribution of the beam intensity after 1.5 m of propagation in GRIN MMF. Inset: beam intensity at the time sample of peak power, showing the process of spatial self-cleaning.

In conclusion, we have developed a new multidimensional characterization technique based on a 3D ST mapping, which allows for capturing, by cumulating several readings, a time-resolved beam image with a resolution of 8 ps. Our method, which is not single-shot, allows for accessing the real-time dynamics of repeatable pulse phenomena. By implementing the 3D ST mapping method to investigate beam propagation in MMFs, we provided the first direct observation of nonlinear mode coupling and cyclic energy exchange between the fundamental mode and the HOMs of a GRIN fiber, thus revealing the complexity associated with beam self-cleaning effect. The role of the SRS process, whose threshold power is larger than that for Kerr beam self-cleaning, was also elucidated. Our results provide a deeper insight in multimode guided wave systems, allowing in particular for extending the characterization of real-time events, which permits a deeper understanding of the physics behind the processes of beam self-cleaning. We believe that our method can help to study several other spatiotemporal processes and interactions, which remained so far largely unexplored, thus possibly significantly advancing their comprehension.

**FUNDING**

H2020 European Research Council (740355); H2020 Marie Skłodowska-Curie Actions (713694); Agence Nationale de la Recherche (ANR-10-LABX-0074-01, ANR-15-IDEX-0003, ANR-16-CE08-0031, ANR-18-CE080016-01); Région Nouvelle Aquitaine (F2MH, SI2P and FLOWA); Direction Générale de l'Armement; iXcore; CILAS (X-LAS); Ministry of Education and Science of the Russian Federation (14.Y26.31.0017).